# Olmec mirrors: an example of archaeological American mirrors

José J. Lunazzi

Universidade Estadual de Campinas - Instituto de Física

13083-970 - Campinas - SP - Brazil

lunazzi@ifi.unicamp.br

## ABSTRACT

Archaeological mirrors from the Olmec civilization are described according to bibliographic references and to personal observations and photographs.

## CONTENTS

# 1. INTRODUCTION

This report was not intended to give all the available information on the subject, but just a simple description that may be valuable for improving the knowledge that the optical community may have on it. The author believes to have consulted most of the available scientific bibliography as it can be traced through cross-referencing from the most recent papers.

Olmec mirrors are the most ancient archaeological mirrors from Mexico and constitute a very good example of ancient American mirrors. The oldest mirrors found in America are from the Incas, made about 800 years before the Olmecs, dated from findings in archaeological sites in Peru. How this technology would have been extended to the north, appearing within the Olmecs, later within the Teotihuacan civilization, a few centuries before the Spanish colonization, is an interesting matter. Mirrors are important also within the Aztec civilization, that appeared in the proximity of the Olmec and Teotihuacan domains at about the time of their extintion.

The extension of the geographic area where these mirrors were employed seems to us not entirely well-known. Those made of Pyrite, the material of highest reflectivity, deteriorate very easily, making it difficult to recognize them as mirrors. Besides this inconvenience, for civilizations living in the jungle, the extension and difficulties of the territory make it hard to obtain archaeological material. More mirrors are expected to be found in future excavations and this may help to better understanding their utilization. Their existance can be associated to the skill of polishing stones with good quality and sphericity. A fundamental property of the iron ore minerals employed is the good reflectivity they have when thoroughly polished.

Most available Olmec mirrors were obtained at sites discovered during this century and studied by some authors. Many of the specimens are merely reported as belonging to unidentified "Private Collections". In many cases the archaeological context of their origin is unknown. They were dated by the radiocarbon method as being between 3125 years up to 2130 years old. Their use, as some archaeologists suggest, goes from fire-making, self-contemplation, medicine, divination, to imaging and astronomy. The range of focal lengths goes from 5cm to more than 80cm for concave surfaces, while convex mirrors, although rare, are known.

Examples of their appearance in iconography are well known and their meaning is the subject of interesting studies. The knowledge the makers had of the laws of reflection and imaging is a very intriguing subject.

# 2. APPEARANCE OF THE MIRRORS

Mirrors are widely spread in South America, the oldest being found at Huaca de los Reyes and Gramalote, on the Peruvian coast[1] and in the Andes at Shillacoto and Kotosh[2]. They also existed in a later period at Chavin de Huántar, contemporary to the Olmecs. At later times in the Andes, polished copper and bronze mirrors were employed by the Incas. In Central America they appeared at a high status burial ground in Sitio Conte on the Rio Grande de Coclé in Panamá.

One of the first general descriptions reported was published in 1926[3], but most excavations have occured since 1940 in a region to the south of Mexico City named La Venta, in the Tabasco state, constituting at present the main site for Olmec mirrors. La Venta, San Lorenzo, Guerrero, Michoacan, Nayarit, Las Bocas - Puebla, San José Mogote - Oaxaca, Chalcatzingo - Morelos, are Mexican sites where Olmec mirrors were found, while Palenque is reported as an example of a site with mirrors from the Mayas. In many cases, the origin of the material and technology employed could not be identified, from which it can be assumed that mirrors were imported from other places. The use of broken mirrors in some offerings has been suggested as an indication that the technology could have disappeared, making the existing elements even more valuable at that time.

### 3. HOW TO FIND THEM

Although two mirrors are reported at the collection of the American Museum of Natural History in United States and some at museums in Europe (British Museum, "Musée de l'Homme"), maybe the best way of seeing them is at the "Museo de Antropologia" in Mexico City. On my first visit to it, I could see at the Pre-Classic section two small feminine ceramic figurines with brilliant mirrors at their breast (Fig.1).

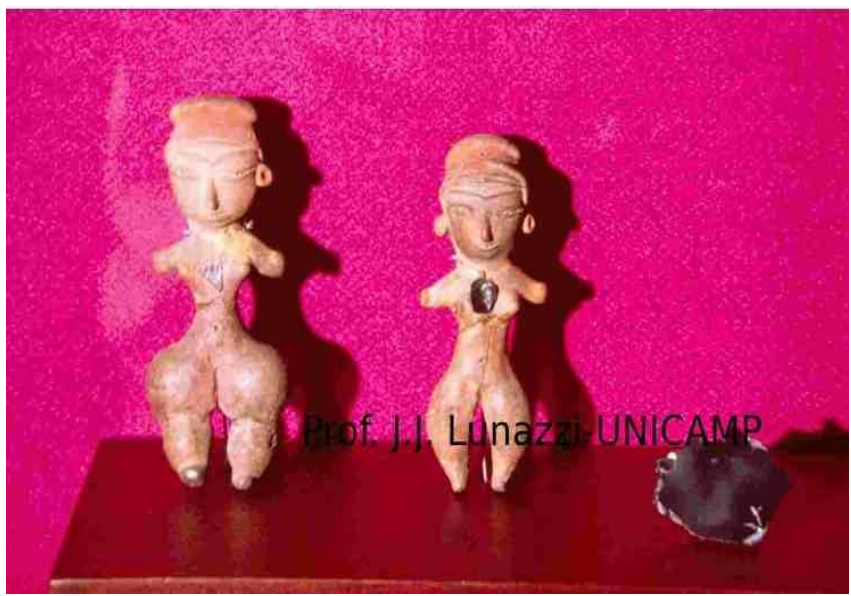

*Fig.1*: *Two feminine Olmec figurines using mirrors, and a mirror sample close to them. Photograph by the author.*

They are referred to by Carlson[4] as findings of the Formative site of Tlatilco, and close to them there is one stone of irregular shape, smaller than 5cm, but very well polished. The observer's face can be seen very clearly reflected, demonstrating the high quality and convexity of the surface[5]. This situation is shown in Fig.2.

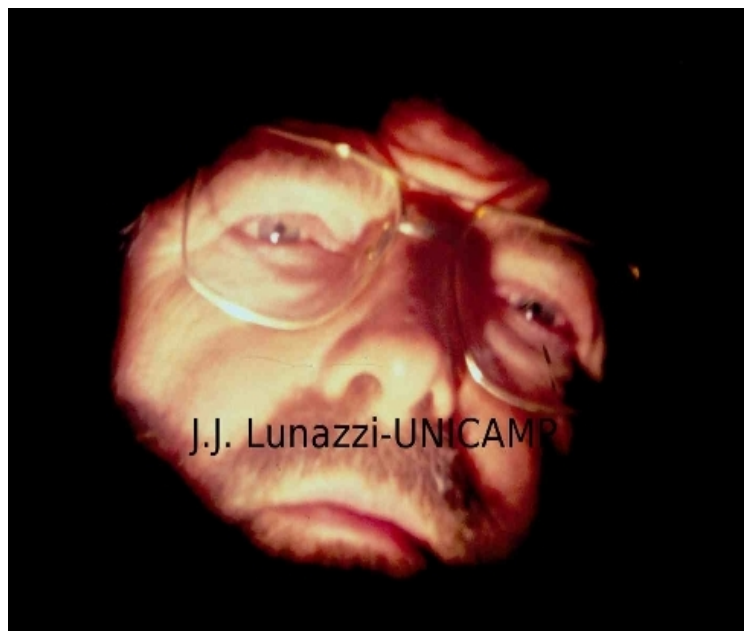

***Fig.2***: *Reflected image of a human face in a convex Olmec mirror. Photograph by the author.*

Its possible use for self-contemplation becomes evident, constituting a small, very portable and strong element. When I saw this unexpected example of archaeological optical techniques, I went excitedly to ask archaeologists for more information. I was told then that there were more Olmec mirrors at the "Costa del Golfo" section. They belong to the findings at La Venta and are concave mirrors laying horizontally on the shelf, but there is one inclined toward the ceiling. In September of 1992, there was a lamp at the ceiling located in a position that made it possible to see its image floating in front of that mirror. Once again, the quality of the image is astonishing. We returned in 1995 to produce the image of a hand that is shown in this article. Six mirrors, five being concave and one convex, can be seen at the permanent exhibit of the "Museo".

## 4. THEIR SIGNIFICANCE IN THE CULTURAL CONTEXT OF THE OLMECS

The iconography shows that mirrors had an important presence in Pre-Columbian civilizations, giving even the name to an Aztec divinity: "Tezcatlipoca", which means "The Smoking Mirror". This god is represented as having a smoking mirror instead of one of his feet. According to a transcription[6], there is an Aztec myth that says that Tezcatlipoca, one of the minor deities, proposed to the other deities to visit the main god Quetzalcoatl, bringing a mirror draped in cotton. When Quetzalcoatl saw his face on the mirror, he cried because he thought himself to be a god but, having a human face, his destination would also be human. Then he left the country announcing his future return. The remaining part of this myth has to do with the belief that the arrival of the Spanish conquerors represented the returning god, but it suffices for us to know that mirrors were an important element on the Aztec mythology. Those mirrors were certainly a legacy from the Olmecs.

In many representations the mirrors are related to the sun god being fixed to his forehead[4] and with internal curved lines. This inner figure was interpreted as the curvature of straight lines distorted by an oblique angle of viewing. It appears displaced, occupying different positions in a way that makes us think about the image of the sun being visible on the mirror. The very well-known sculptures of a feathered serpent, very common at the Teotihuacan site, represent the serpent traversing the mirror[7]. To give an example of the capability of generating real images, we made the photograph of the image of one hand reflected in a concave Olmec mirror (Fig.3).

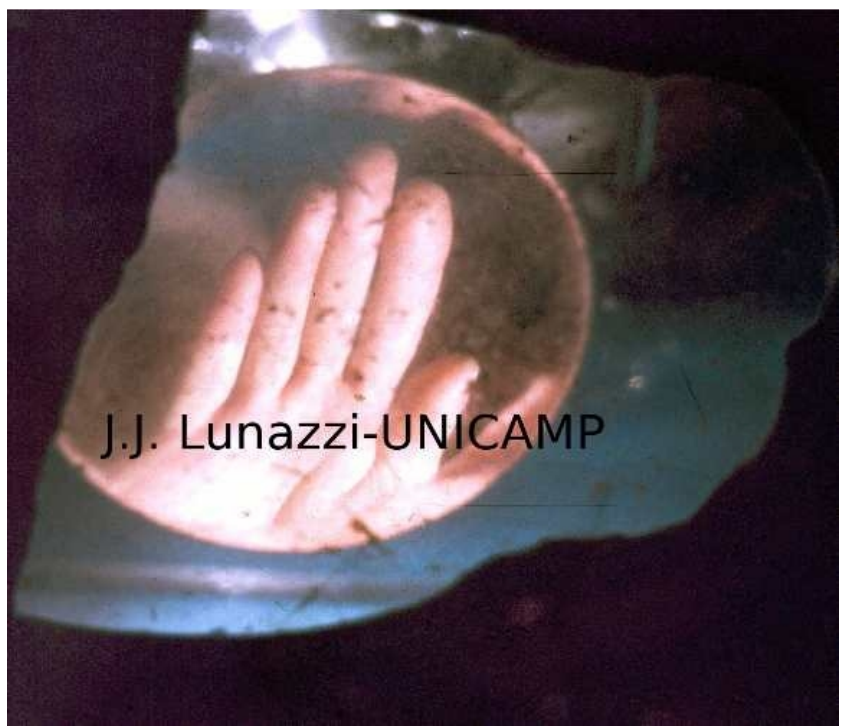

**Fig.3**: *Image of a hand obtained by means of a concave Olmec mirror. Photograph by the author.*

These are good images although we could not work at ideal conditions: the focal length of a 105mm objective was doubled by means of a diverging additional lens, and we were not allowed to remove the protective glass, which was traversed twice by the light at an angle of about 70°. We employed an aperture for the diaphragm of the objective on the camera of about 5mm, to obtain an image similar to the one obtained by the naked eye. We employed KODAK TMAX 400 ISO film. Although there was ordinary glass between the mirror and the camera, and some deterioration may have occured on the mirror surface, changing its original conditions, the image is very clear and appears floating in the air in front of the mirror. None of the bibliographic references we consulted indicate the possibility of images from the mirrors being present on the iconography as inverted images, which would certainly be an interesting finding.

We can imagine what strong feelings these images would have caused on the Olmec people from the registers existing of other close civilizations also. Drawings associating mirrors to serpents, human faces, eyes, cotton, water, flowers, butterflies and others were described[7] for the Teotihuacan civilization. The representation of mirrors with a human-like eye inside could be the consequence of directly looking into a convex mirror of short focal length, but it is also known that looking at mirrors was interpreted as getting in contact with a magic domain. Mirrors were associated also to the jaguar's eyes because of the reflections that can be seen on them. In many native languages, the Amerindians have the same or very close words for designating "mirror" and "eye". Regarding the Aztecs, for example, in Book 10 of the Florentine Codex, both the eye and the pupil are described as "tezcatl" (mirror) [7]. The word for mirror at the central Maya domain was "nen" derived from "lem", which means something bright, gleaming and reflective. In Yucatec phrases, "nen" denotes rulers or persons of preeminent social status as "the reflection of the world" [8].

Mica and stones were used to represent the eyes of divinities in Teotihuacan sculptures, while other reflective materials, such as mercury or pearls could have had similar importance in relating their properties to spiritual worlds.

Making sacred fire by means of the sun is the most probable use of some mirrors, mainly for those which are spherical, although we know very little about successful experiences of firing with the original mirrors. The association of mirrors with cotton[7] was suggested by us[5] as a possible consequence of the Inca technique of making fire with mirrors described by the Inca Garcilaso[9] in the 16th century since it would be a way of obtaining more efficiency for starting fire than with dry wood or leaves .

When fire was obtained in this way in ritual sacrifices, this was interpreted as being the people in the grace of the gods, as reported by Garcilaso. Mirrors were used by the priests and the nobles, as a pectoral or on the back, probably being a symbol of high social status. We can get this impression from the photograph at Fig.4.

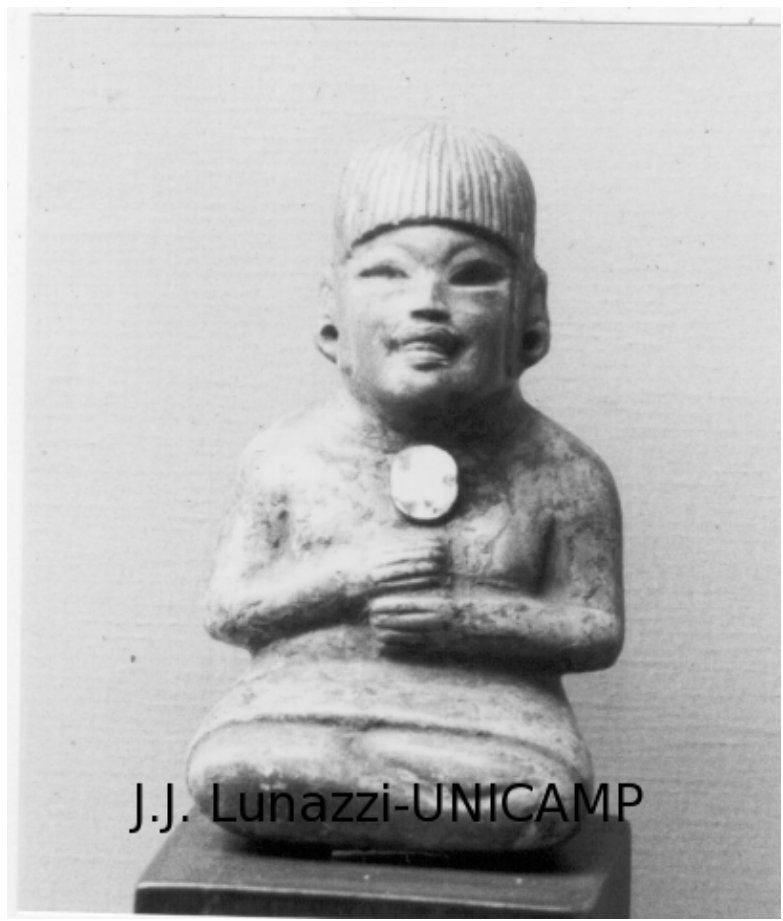

**Fig.4** *: Olmec representation where a reflecting element is located at the chest. Jade figurine, 8cm high. Photograph by the author.*

It has been suggested that the Olmec elite identified themselves with a mythical jaguar ancestor. Even now some Amerindians preserve this identification, dressing as man-jaguars with a complete jaguar hide. One Olmec monument shows a jaguar-man figure apparently copulating with a human female, and a possible mirror pendant is seen suspended on the chest of the jaguar-like figure. The use as a pectoral is seen not only in the iconography but at the chest position of skeletons at burials.

The most probable uses of mirrors are: divination, introspection, firing, imaging and medicine. Real images can be observed not only floating in the air but also projected on a white, partially covered surface[10]. It was mentioned that in the Chinese culture, a 1.000 year old technique was reported where mirrors were applied for cauterizing under the sun light. Plane mirrors are not reported for the Olmecs, which is an intriguing fact due to the utility they could have in light communications.

I would like to mention that many of us were taught at school that when the Spanish conquerors came to America they exchanged mirrors for many valuables and goods with the natives. This led us to think that they were marveled by the images on the mirrors, which must be true, but also that they were not familiar with those images. Now we can think of a different history; that the mirror images were considered to be divine or highly important. Then the people were taught to obey the priests and authorities who carried mirrors within their vestments. A natural transference of obeisance could had happened when seeing the mirrors the newcomers brought with them.

## 5. TYPES OF MIRRORS

Mirrors made in a single piece were ground mainly from iron ore minerals, which have the remarkable property of good reflectivity when very well polished. The reflectivity of the mirrors is not reported in the references but the same materials have reported reflectivities of 21% for Magnetite, 28% for Hematite, 55% for Pyrite[11].

Mica and Obsidian are low-reflecting materials also employed. In the case of Pyrite, the presence of 0.1% of gold within its structure was demonstrated, not visible at the optical microscope. Being the most brilliant within the iron-ore materials, it is the easiest to deteriorate under humid conditions, to the point of leaving just a residual deposit. Pyrite seems not to have being available to the Olmecs but reported to be found within the Mayas. The grinding materials could have been any of the elements available at that time, such as emery (aluminum oxide), hematite powder, or maybe sand. Ochre (hematite jeweler's rouge) was used for polishing, a very patient task. These materials are not different from the ones employed now-a-days for making mirrors on glass. It has been indicated that the polish of the specimens is so good that it represents the limit of perfection that the material will allow[10]. No trace of abrasion marks was found and the microstructure of the mineral was revealed. This result can only be obtained in modern technology by combining polishing and etching. When grinding, sphericity is a consequence of the lapidary process. A pair of complementary concave-convex surfaces is obtained, but it is clear that the preference was given to the concaves because convex surfaces were reshaped into concave. It may be also an indication of not having too much material available. Minute, local irregularities in the curvature of the surfaces suggest that the work was done on small areas at a time. In order to better analyze the procedure of fabrication, a site at San José Mogote - Oaxaca, is mentioned as a probable artisan workshop area, deserving further study.

A great variety of curvatures can be found, generating focal lengths from -50cm [5] to more than 80cm. It is probable that the shorter focal lengths (about 5cm) were the most appropriate for burning purposes. The existence of many mirrors whose vertical and horizontal curvature do not coincide may be an indication of a different use, not precluding that of imaging[5]. A parabolization of the surface intended to better concentrate the light is mentioned[10], although not characterized in detail. Plane mirrors are not explicitily referenced but it seems very probable that they existed. The author could see one, reported as belonging to a private collection, and also received a suggestion of its possible use in making signals by lighting remotely toward the observer.

Regarding the diameters of the mirrors, they corresponded to circular or elliptical shapes, the maximum reported diameter being 14cm. Mosaic mirrors and the use of Pyrite are not reported for the Olmecs but for the Teotihuacan and Kaminaljuyu findings. They correspond to civilizations appearing in the Mexican and Guatemalan regions, later than the Olmecs. For them, the assembling of pieces may have allowed mirror diameters of up to 50cm. No good images were reported for mosaic mirrors.

The quality of Olmec mirrors, although variable, can be very high, as it can be seen from the images they generate at the short distance of a few millimeters[12] or even at the large distance of some meters[5]. Observation at the microscope showed good quality because no scratches were visible, although smoothness, roughness, reflectivity or aberration coefficients of the surfaces were not reported.

## 6. ON THE QUALITY OF REFRACTIVE ELEMENTS

Very well polished quartz objects are known to belong to the Olmec culture. Among them there is a small sphere, smaller than 10mm in diameter, within the group of offerings of the E tomb at La Venta, Tabasco, in the permanent exhibit of the "Museo Nacional de Antropologia" at Mexico City (Fig.5, second from left on the upper line).

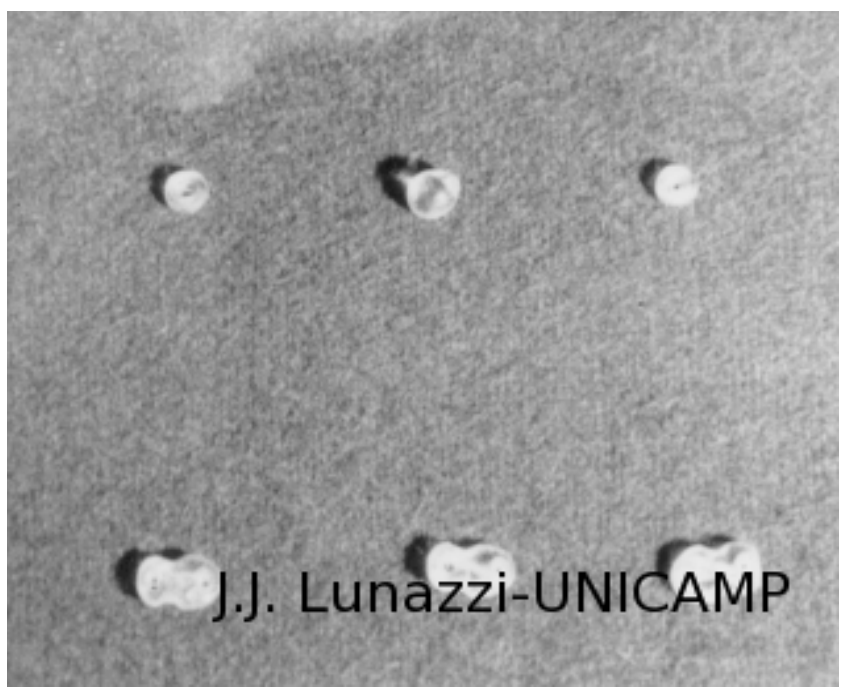

**Fig.5:** *Some small well-polished Olmec objects made of quartz, including one sphere. Photograph by the author.*

Although we could not have the opportunity of seeing an object through it, its appearance made us think that this element may function as a magnifying lens and that the knowledge the Olmecs had about optics could include some refractive properties. The focusing of the illuminating lamp is clear at the upper left side of the sphere.

## 7. CONCLUSIONS

From the information we obtained we can recognize the Olmec mirrors as very valuable cultural elements of unusual quality for ancient cultures. The precise characterization of their optical parameters seems not to have been completely reported, so that its measurement and analysis can still be an interesting subject of research in an interdisciplinary field involving archaeology and optics in close relationship.

## 8. ACKNOWLEDGMENTS

The author wish to express his acknowledgment to the Organizing Committee of the "II Reunión Iberoamericana de Óptica" , Guanajuato - GTO, Mexico, 18-22 September 1995 for their invitation to present this work at that meeting, particularly to its General President Dr. Daniel Malacara Hernandez whose broad way of thinking and acting has allowed the realization of many important activities in Latin America.

Financial support from FAEP - Campinas State University, FAPESP - Foundation for Assistance to Research of the Sao Paulo State, CNPq - National Council of Research made this work possible. I am also grateful to Jesus Najera who, representing the Esperanto world community, helped with equipment and personal work to obtain the photographs. The "Museo Nacional de Antropologia" of Mexico City is acknowledged for allowing us to introduce the photographic and video cameras, their mechanical holders and a lamp into their exhibiting hall to make the photographs. Archaeologist Marcia L. Castro is acknowledged for the information reported at that museum in October 1992. Alexsandra Siqueira is acknowledged for helping at the photographic laboratory.

**Color Plate printed at the center of the book**
**(it is inverted right-left in this reproduction):**

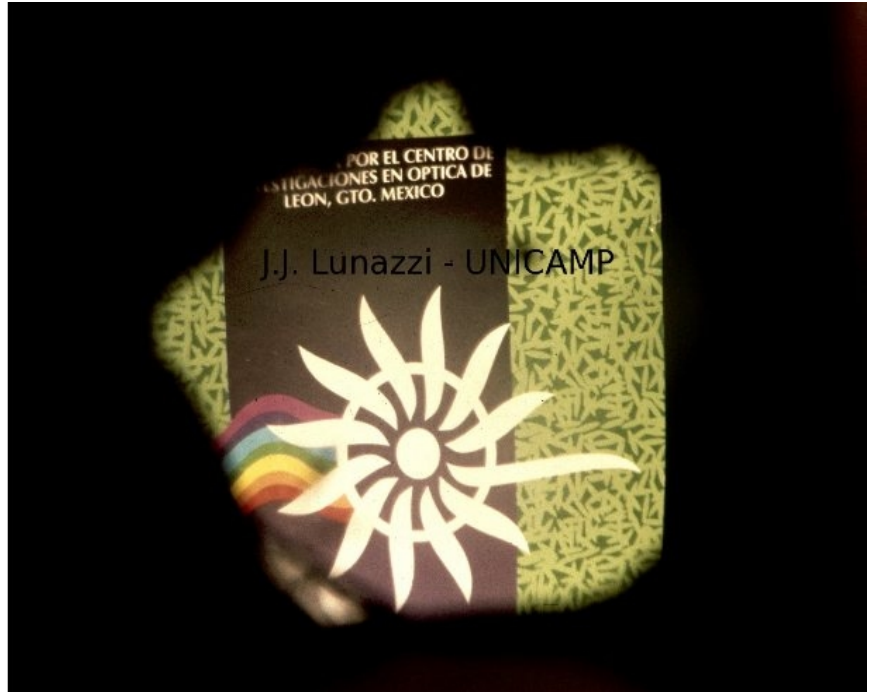

*Plate 22.1. Image of printed letters and colour symbol reflected from an Olmec mirror. Photographed by the author.*